\documentclass[preprint,prc,superscriptaddress,unsortedaddress,showpacs,preprintnumbers,amsmath,amssymb]{revtex4}
\usepackage{tipa}

\usepackage{graphicx}
\usepackage{dcolumn}
\usepackage{bm}

\def\beq{\begin{equation}}
\def\eeq{\end{equation}}
\def\bea{\begin{eqnarray}}
\def\eea{\end{eqnarray}}

\def\fun#1#2{\lower3.6pt\vbox{\baselineskip0pt\lineskip.9pt
  \ialign{$\mathsurround=0pt#1\hfil##\hfil$\crcr#2\crcr\sim\crcr}}}

\preprint{}

\begin{document}

\title{Constraints on neutron skin thickness in $^{208}$Pb and density-dependent symmetry energy}

\author{Jianmin Dong}\email[ ]{dongjm07@impcas.ac.cn}\affiliation{Institute of Modern Physics, Chinese
Academy of Sciences, Lanzhou 730000, China}
\author{Wei Zuo}\email[ ]{zuowei@impcas.ac.cn}
\affiliation{Institute of Modern Physics, Chinese Academy of
Sciences, Lanzhou 730000, China}
\author{Jianzhong Gu}\email[ ]{jzgu1963@ciae.ac.cn}
\affiliation{China Institute of Atomic Energy, P. O. Box 275(10),
Beijing 102413, China}

\begin{abstract}
Accurate knowledge about the neutron skin thickness $\Delta R_{np}$
in $^{208}$Pb has far-reaching implications for different
communities of nuclear physics and astrophysics. Yet, the novel Lead
Radius Experiment (PREX) did not yield stringent constraint on the
$\Delta R_{np}$ recently. We employ a more practicable strategy
currently to probe the neutron skin thickness of $^{208}$Pb based on
a high linear correlation between the $\Delta R_{np}$ and
$J-a_{\text{sym}}$, where $J$ and $a_{\text{sym}}$ are the symmetry
energy (coefficient) of nuclear matter at saturation density and of
$^{208}$Pb. An accurate $J-a_{\text{sym}}$ thus places a strong
constraint on the $\Delta R_{np}$. Compared with the
parity-violating asymmetry $A_{\text{PV}}$ in the PREX, the reliably
experimental information on the $J-a_{\text{sym}}$ is much more
easily available attributed to a wealth of measured data on nuclear
masses and on decay energies. The density dependence of the symmetry energy is also
well constrained with the $J-a_{\text{sym}}$. Finally, with a `tomoscan' method, we find that
one just needs to measure the nucleon densities in $^{208}$Pb
starting from $R_{m} = 7.61\pm0.04$ fm to obtain the $\Delta R_{np}$
in hadron scattering experiments, regardless of its interior profile
that is hampered by the strong absorption.
\end{abstract}

\maketitle

\section{Introduction}\label{intro}\noindent
 The nuclear physics overlaps and interacts with astrophysics not
only expands its research space but also promotes the development of
fundamental physics. A great of attention has been paid to the
equation of state (EOS) of isospin asymmetric nuclear matter in both
the two fields as the development of radioactive beam facilities and
astronomical observation facilities over the past decade. The
symmetry energy that characterizes the isospin dependence of the
EOS, is a quantity of critical importance due to its many-sided
roles in nuclear physics \cite{VB,BAL,PD,AWS,mass1,mass2,JD} and
astrophysics \cite{HTJ,NS,KH,flare,Wen,QSO,LF}. Although great
efforts have been made and considerable progresses have been
achieved both theoretically and experimentally, its density
dependence ultimately remains unsolved because of the incomplete
knowledge of the nuclear force as well as the complexity of
many-body systems. Nevertheless, many important and leading issues
in nuclear astrophysics require the accurate knowledge about it
ungently at present.

The symmetry energy $S(\rho )$ of nuclear matter is usually expanded
around saturation density $\rho_{0}$ as
\begin{equation}
S(\rho )=J + \frac{L}{3}\left( \frac{\rho -\rho _{0}}{\rho _{0}}\right) +%
\frac{K_{\text{sym}}}{18}\left( \frac{\rho -\rho _{0}}{\rho
_{0}}\right) ^{2}+...,\label{A}
\end{equation}
where $J=S(\rho_{0} )$ is the symmetry energy at $\rho_{0}$. The
slope parameter $L=3\rho
\partial S(\rho )/\partial \rho |_{\rho _{0}}$ and curvature parameter
$K_{\text{sym}}=9\rho ^{2}\partial^{2}S/\partial \rho ^{2}|_{\rho
_{0}}$ characterize the density-dependent behavior of the symmetry
energy around $\rho_{0}$. Extensive independent studies have been
performed to constrain the slope $L$, but the uncertainty is still
large \cite{MBT2,XV,BMS,BAL1}.

It has been established that the slope parameter $L$ is strongly
correlated linearly with the neutron skin thickness $\Delta R_{np}$
of heavy nuclei \cite{BAB,ST,RJF}. Although the theoretical
predictions on $L$ and $\Delta R_{np}$ are extremely diverse, this
linear correlation is universal in the realm of widely different
mean-field models \cite{XRM}. Accordingly, a measurement of $\Delta
R_{np}$ with a high accuracy is of enormous significance to
constrain the density-dependent behavior of $S(\rho )$ around
$\rho_{0}$. Actually, many experimentalists have been concentrating
on it with different methods including the x-ray cascade of
antiprotonic atoms \cite{EA}, pygmy dipole resonance \cite{PDR,AK},
proton elastic scattering \cite{JZ}, proton inelastic scattering
\cite{EDR} and electric dipole polarizability \cite{JBG}. However,
systematic uncertainties associated with various model assumptions
are unavoidable. The parity-violating electron elastic scattering
measurement in the parity radius experiment (PREX) at the Jefferson
Laboratory combined with the fact that the parity-violating
asymmetry $A_{\text{PV}}$ is strongly correlated with the neutron
rms radius, determined the $\Delta R_{np}$ to be
$0.33^{+0.16}_{-0.18}$ fm with a large central value compared to
other measurements and analyses \cite{SA}. Although it was suggested
that ruling out a thick neutron skin in $^{208}$Pb seems premature
\cite{FJF}, in any case, the large uncertainty seems to be not of
much help to explore the symmetry energy and other interesting
issues. In this work, a more practicable strategy compared with
the PREX at current is introduced to probe the $\Delta R_{np}$ of
$^{208}$Pb together with nuclear matter symmetry energy. A new
insight into the neutron skin is also provided.

\section{Neutron skin thickness $\Delta R_{np}$ probed by the $J-a_{\text{sym}}$}\label{intro}\noindent
The neutron skin thickness of nuclei is given as $\Delta R_{np}=\sqrt{\frac{3}{5}}\left[ \frac{2r_{0}}{3J}%
(J-a_{\text{sym}}(A))A^{1/3}(I-I_{c})-e^{2}Z/(70J)\right] +S_{sw}$
in the nuclear droplet model \cite{DM1,DM2} with isospin asymmetry $I$,
nuclear radius $R=r_{0}A^{1/3}$ and a correction
$I_{c}=e^{2}Z/(20JR)$ due to the Coulomb interaction. $Z$, $A$ are
the proton and mass numbers, respectively. $S_{sw}$ is a correction
caused by an eventual difference in the surface widths of nucleon
density profiles. $a_{\text{sym}}(A)$ is symmetry
energy (coefficient) that has been received
great interest because with the help of it one may obtain
some information on the density dependence of $S(\rho )$ \cite{Dong20131,Tian,Shen1}. Centelles {\it et al}. showed that the neutron
skin thickness $\Delta R_{np}$ correlates linearly with
$J-a_{\text{sym}}(A)$ based on different mean-field models, where
the symmetry energy (coefficient) $a_{\text{sym}}(A)$ is obtained
within the asymmetric semi-infinite nuclear matter (ASINM)
calculations \cite{DM2}. In our previous work, instead of using the
ASINM calculations, the $a_{\text{sym}}(A)$ was obtained in the
framework of the Skyrme energy-density functional approach by
directly integrating the density functional of the symmetry energy
after subtracting Coulomb polarization effect without introducing additional
assumptions \cite{Dong20131}. In the present work, the
$a_{\text{sym}}(A)$ of $^{208}$Pb, marked as $a_{\text{sym}}$, is
extracted with both the Skyrme effective interactions and
relativistic effective interaction Lagrangians, and the local
density approximation is adopted by dropping the negligible
non-local terms compared to \cite{Dong20131}. As done in Ref.
\cite{XRM}, to prevent eventual biases, we avoid including more than
two models of the same kind fitted by the same authors and protocol
and avoid models providing a charge radius of $^{208}$Pb away from
experiment data by more than $1\%$.

\begin{figure}[htbp]
\begin{center}
\includegraphics[width=0.65\textwidth]{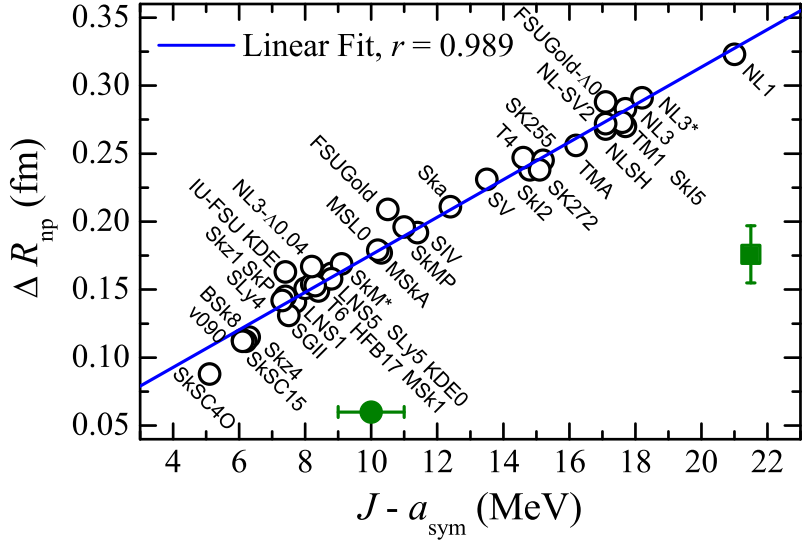}
\caption{(Color Online) Neutron skin thickness $\Delta R_{np}$ in $^{208}$Pb
against the $J-a_{\text{sym}}$ with different nuclear energy-density
functionals.}
\end{center}
\end{figure}

The calculated neutron skin thickness $\Delta R_{np}$ of $^{208}$Pb
and $J-a_{\text{sym}}$ with different mean-field models are
presented in Fig. 1, in which a close dependence of $\Delta R_{np}$
on $J-a_{\text{sym}}$ predicted by the droplet model is displayed.
By performing a two-parameter fitting, the correlation is given by
\begin{equation}
\Delta
R_{np}=(0.0138\pm0.0003)(J-a_{\text{sym}})+(0.0376\mp0.0041),\label{B}
\end{equation}
with the correlation coefficient $r=0.989$, where $\Delta R_{np}$
and $J-a_{\text{sym}}$ are in units of fm and MeV, respectively.
Here the empirical saturation density $\rho_{0}=0.16$ fm$^{-3}$ \cite{book} is
used uniformly. If the symmetry energy is calculated at their own
saturation densities from the mean-field models, the linear
correlation vanishes due to the fact that the relativistic
interactions provide smaller saturation densities compared with the
non-relativistic ones. The $\Delta R_{np}$ of $^{208}$Pb is
found to have a high linear correlation with $J-a_{\text{sym}}$
as that with the slope $L$ (not shown here). It is thus
indisputable that the $J-a_{\text{sym}}$ with a high accuracy places
a stringent constrain on the $\Delta R_{np}$. As the primary
advantage, the reliably experimental information about the
$J-a_{\text{sym}}$ is much more easily available compared with that
about the parity-violating asymmetry $A_{\text{PV}}$ in the PREX.
Recently, the symmetry energy $J$ at saturation density $\rho_{0}$
has been well determined to rather narrow regions, in particular,
$32.5\pm0.5$ MeV from the mass systematics \cite{FRDM} and $32.10\pm
0.31$ MeV from the double differences of ¡°experimental¡± symmetry
energies \cite{HJ} agreeing with that of the mass systematics. These
results are very useful in exploring the density-dependent symmetry
energy as inputs \cite{BKA}. Here we adopt the union of the two
values, i.e. $J=32.4\pm0.6$ MeV, and hence the central issue is to
determine the symmetry energy $a_{\text{sym}}$ of $^{208}$Pb
accurately. We extract the mass dependent symmetry energy
$a_{\text{sym}}(A)=J/(1+\kappa A^{1/3}) $
 \cite{DROP1,DL} with $\beta^{-}$-decay energies $Q_{\beta ^{-}}$ of
heavy odd-$A$ nuclei and with mass differences $\Delta B$ between
$^{A}(Z-1)$ and $^{A}(Z+1)$ as our previous calculations
\cite{dong13,dong14} but with a new input quantity $J$, and then
derive the $a_{\text{sym}}$ of $^{208}$Pb. The merit of these two
approaches is that only the well known Coulomb energy survives in
$Q_{\beta ^{-}}$ and in $\Delta B$ when determining the unknown
$a_{\text{sym}}$, where the $Q_{\beta ^{-}}$ and $\Delta B$ are all
taken from experimental data. Consequently, the $a_{\text{sym}}$ is
extracted to be $22.4\pm0.4$ MeV accurately, which is quite
insensitive to the input quantity $J$. As a result, the derived
$J-a_{\text{sym}}$ is $10.0\pm1.0$ MeV (solid circle in Fig. 1), which allows us to constrain
the neutron skin thickness as well as the slope $L$ in our
subsequent calculations.

The neutron skin thickness in $^{208}$Pb is predicted to be $\Delta
R_{np}=0.176\pm0.021$ fm (solid square in Fig. 1), where the estimated error stems from the
uncertainties of the $J-a_{\text{sym}}$ as well as Eq. (\ref{B}). To
reach such an error level, the $A_{\text{PV}}$ in the PREX should be
measured at least up to $2\%$ accuracy, which is hardly implemented at
present. This fact indicates the $J-a_{\text{sym}}$ is much more
effective to probe the $\Delta R_{np}$ currently. The precise
information about the $\Delta R_{np}$ is of fundamental importance
and has far-reaching implications in neutron star physics, such as
the structure, composition and cooling. As an example, a relation of
$\rho_{c}\approx0.16-0.39\Delta R_{np}$ was put forward to describe
the relation between the $\Delta R_{np}$ of $^{208}$Pb and the
transition density $\rho_{c}$ from a solid neutron star crust to the
liquid interior \cite{NS1}, where the $\rho_{c}$ is estimated to be
$0.091\pm0.008$ fm$^{-3}$. The properties of the crust-core
transition is of crucial importance in understanding of the pulsar
glitch \cite{glitch}.

\section{Density dependence of the symmetry energy probed by the $J-a_{\text{sym}}$}\label{intro}\noindent
Since the neutron skin thickness $\Delta R_{np}$ correlates linearly with both the
slope $L$ and $J-a_{\text{sym}}$, the slope $L$ naturally
correlates linearly with the $J-a_{\text{sym}}$, which is displayed
in Fig. 2(a). The linear relation is $L=(9.682\pm0.285)(J-a_{\text{sym}})+(-42.694\mp3.441)$, where $L$
and $J-a_{\text{sym}}$ are in units of MeV. Imposing the above obtained
$J-a_{\text{sym}}$, the slope parameter is estimated to be
$L=54\pm16$ MeV. Recently, the properties of nuclear matter at
subsaturation density $\rho \approx 0.11$fm$^{-3}$ have attracted
considerable attention because it has been shown that the $\Delta
R_{np}$ is uniquely fixed by the slope $L(\rho \approx 0.11$
fm$^{-3})$ \cite{chen1} and the giant monopole resonance of heavy
nuclei is constrained by the nuclear matter EOS at this density
\cite{EMV}. Fig. 2(b) shows that the slope
$L(\rho=0.11\text{fm}^{-3})$ (labeled $L_{0.11}$ for short) and
$J-a_{\text{sym}}$ have a higher linear dependence
$L_{0.11}=(4.542\pm0.073)(J-a_{\text{sym}})+(2.140\mp0.885)$ with
the correlation coefficient $r=0.995$. Accordingly, the $L_{0.11}$
is evaluated to be $48\pm 6$ MeV.

\begin{figure}[htbp]
\begin{center}
\includegraphics[width=0.65\textwidth]{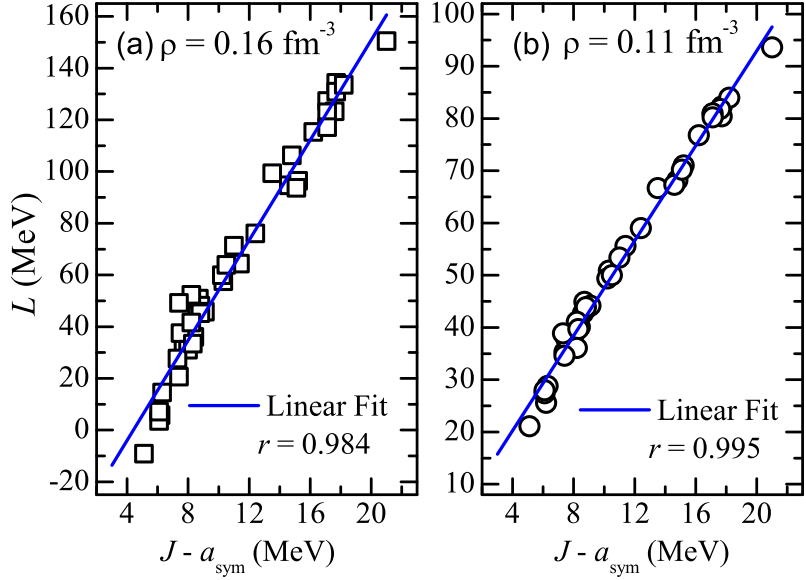}
\caption{(Color Online) Correlation of the slope parameter $L$ at densities
$\rho=0.16$ fm$^{-3}$ and $\rho=0.11$ fm$^{-3}$ with the
$J-a_{\text{sym}}$.}
\end{center}
\end{figure}

The slope $L$ is constrained with the $J-a_{\text{sym}}$ in another
way for comparison. Centelles {\it et al}. found that the symmetry
energy $a_{\text{sym}}$ of $^{208}$Pb is approximately equal to the
nuclear matter symmetry energy $S(\rho_{A})$ at a reference density
$\rho_{A} \simeq 0.1$ fm$^{-3}$ \cite{DM2}. This important relation
bridges the symmetry energies of nuclear matter and the finite
nucleus. We calculate the reference density $\rho_{A}$ for
$^{208}$Pb and find that the interactions which provide the values
of $J$ and $a_{\text{sym}}$ agreeing with the ones extracted from
experimental information, give $\rho_{A}\simeq 0.088$ fm$^{-3}=0.55
\rho_{0}$. It should be noted that the $a_{\text{sym}}$ does not
equal the symmetry energy at the mean density of $^{208}$Pb as a
result of the extremely inhomogeneous isospin asymmetry distribution
in the nucleus as shown in \cite{Dong20131}. Since the accurate
value of the reference density $\rho_{A}$ is of crucial importance
for determining the slope parameter $L$ \cite{dong13,dong14}, we
further examine it. Instead of the DDM3Y-shape expression used
before \cite{dong13,dong14}, Eq. (\ref{A}) is employed directly to
describe the density dependent symmetry energy to reduce the
uncertain factors as far as possible. The $K_{\text{sym}}$ term
that contributes weakly to the symmetry energy nearby $\rho_{0}$ is estimated with the
relation $K_{\text{sym}}=39+5 L-15 J$ \cite{dong12} obtained from
the DDM3Y-shape expression without loss of accuracy. In terms of
$J-S(\rho_{A})=10.0 \pm 1.0$ MeV and $\rho_{A}=0.55 \rho_{0}$, the
slope $L$ at the saturation density $\rho_{0}$ is predicted to be
$53 \pm 10$ MeV according to Eq. (\ref{A}), which is in excellent
agreement with that from Fig. 2(a). At the density of $\rho=0.11$
fm$^{-3}$, the slope $L_{0.11}=49\pm4$ MeV, being also particularly
consistent with the value of $48\pm6$ MeV from Fig. 2(b). The
consistency of the two approaches not only indicates the reliability
of the present methods but also further verifies the accuracy of the
reference density $\rho_{A}=0.55 \rho_{0}$. As an important
conclusion, the $a_{\text{sym}}=S(\rho=0.55\rho_{0})\simeq
22.4$ MeV will be a very useful reference to calibrate the effective
interactions in nuclear energy density functionals.

With the obtained $L_{0.11}$ and $L$ values, the curvature parameter is
evaluated to be $K_{\text{sym}}=-152\pm70$ MeV. Currently,
the symmetry energy at suprasaturation densities is
extremely controversial. It was indicated that the three bulk
parameters $J$, $L$ and $K_{\text{sym}}$ well characterize the
symmetry energy at densities up to $\sim 2\rho_{0}$ while higher
order terms contribute negligibly small \cite{Chen3}. If true, the symmetry energy
$S({\rho})$ at high densities up to $\sim 2\rho_{0}$ turns out to be
not stiff, as shown in Fig. 3. The symmetry energy at $2\rho_{0}$ is
estimated to be $S(2\rho_{0})=42\pm10$ MeV. In short, to characterize the
symmetry energy at high densities, the accurate knowledge about its
density dependence at the saturation density is crucial.

\begin{figure}[htbp]
\begin{center}
\includegraphics[width=0.65\textwidth]{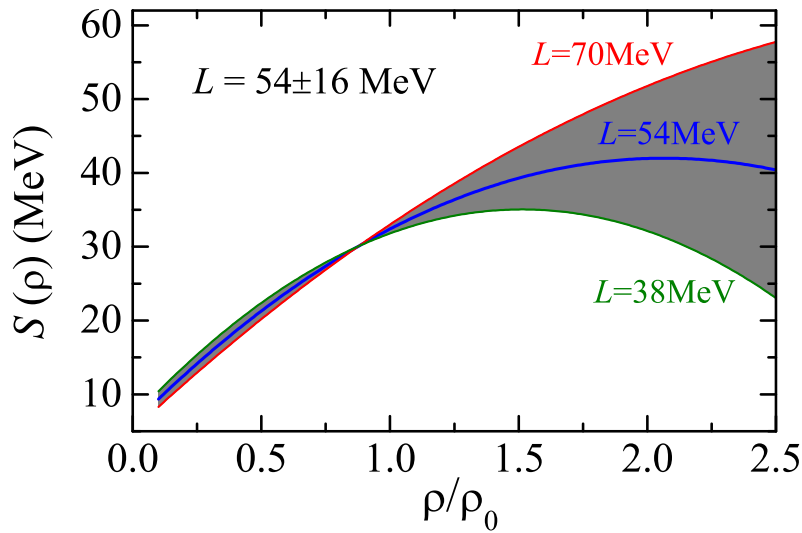}
\caption{(Color Online) Density dependent symmetry energy at high
densities.}
\end{center}
\end{figure}

\section{Further exploration on the measurement of the $\Delta R_{np}$}\label{intro}\noindent
Based on the above discussions on the neutron skin thickness $\Delta R_{np}$ and symmetry energy,
we make an exploration on the measurement of the $\Delta R_{np}$ in
$^{208}$Pb. To grasp richer information on the $\Delta R_{np}$, we
formulate it as an integral of a distribution function
\begin{eqnarray}
\Delta R_{np} &=&\sqrt{<r_{n}^{2}>}-\sqrt{<r_{p}^{2}>}
=\int_{0}^{\infty }f(r)dr,
\end{eqnarray}
where $f(r)=4\pi r^{4}\left( \frac{\rho _{n}}{N}-\frac{\rho
_{p}}{Z}\right) /\left( \sqrt{<r_{n}^{2}>}+\sqrt{<r_{p}^{2}>}\right)
$ is defined as the radial distribution function which is actually
determined by the nucleon densities and reflects the detailed
information about the neutron skin.
$\sqrt{<r_{n}^{2}>}+\sqrt{<r_{p}^{2}>} \simeq 11.1$\ fm changes by
less than $3\%$ in the mean field model calculations, and can be
taken as a known value. Fig. 4(a) illustrates the distribution
function $f(r)$ in $^{208}$Pb as a function of distance $r$
generated by the SLy5 interaction as an example. It is a misleading
idea to consider the neutron skin merely originating
from the nuclear surface. The area enclosed
by the x-axis and the curve $f(r)$ (colored regions) is exactly the
neutron skin thickness $\Delta R_{np}$. We name this new method that
dissects the $\Delta R_{np}$ with a distribution
function as `tomoscan' picturesquely here. As a new concept in nuclear physics, it could also be used to
analyze other intriguing issues, such as the halo structure in
exotic nuclei. The region of $r<R_{0}$ contributes negatively while
that of $r>R_{0}$ contributes positively to the $\Delta R_{np}$.
Thus, there exists a distance $R_{m}$ below which ($0\le r<R_{m}$)
the contributions (red shaded regions) cancel each other out, and hence the
$\Delta R_{np}$ can be calculated by the neutron and proton density
distributions just starting from $R_{m}$ (blue filled region).

\begin{figure}[htbp]
\begin{center}
\includegraphics[width=0.65\textwidth]{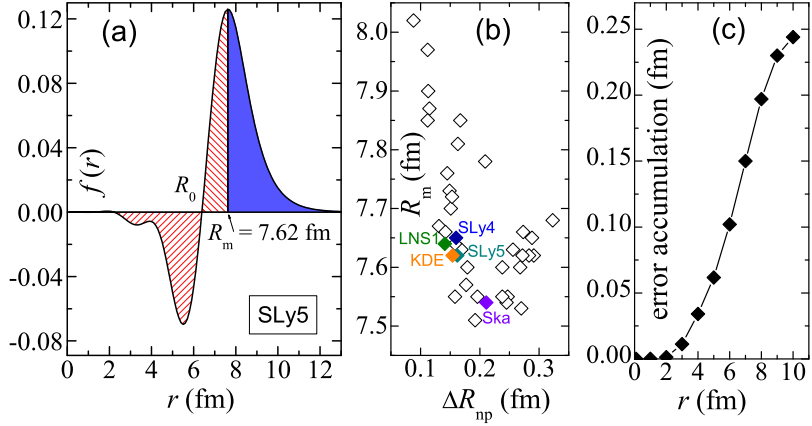}
\caption{(Color Online) (a) Radial distribution function $f(r)$ of
the neutron skin thickness in $^{208}$Pb. The contributions from the two
parts in the red shaded regions cancel each other out. The area under the
curve of $f(r)$ starting from $R_{m}$ (blue filled region) is equal
to the neutron skin thickness $\Delta R_{np}$. (b) Calculated $R_m$
values with different energy density functionals. The colored solid symbols
are from the interactions generating the reference density $\rho_{A}
\simeq 0.55\rho_{0}$, $a_{\text{sym}}\simeq 22.4$ MeV
and $J \simeq 32.4$ MeV. (c) Error accumulation of the $\Delta
R_{np}$ measurement in hadron scattering experiments as a function
of distance $r$, where the nucleonic density distributions are from
Tables III and IV in Ref. \cite{JZ}.}
\end{center}
\end{figure}

The calculated values of $R_{m}$ with different interactions are marked
in Fig. 4(b). The $R_{m}$ is found to be model dependent, which should
be further constrained. The interactions generating
smaller (larger) $\Delta R_{np}$ tend to yield slightly larger (smaller) $R_{m}$.
As we mentioned above, one important conclusion of this work is that
the $a_{\text{sym}}=S(\rho=0.55\rho_{0})\simeq 22.4$ MeV (along with
$J \simeq 32.4$ MeV) serve important calibrations for effective interactions
in nuclear energy density functionals. Thus we use those constraint
conditions to filter those interactions. The eligible interactions
give $R_{m} = 7.61 \pm 0.04 $ fm (colored solid symbols), where the error
bar of $\pm 0.04$ fm just leads to an uncertainty of the $\Delta
R_{np}$ by about $\pm 0.005$ fm. The error bar of $\pm 0.005$ fm for the
$\Delta R_{np}$ is so small that the obtained $R_m$ value should not be
regarded as model dependent any more. This result leads to an intriguing
conclusion: one just needs to measure the rather dilute matter
located in the nuclear surface to determine the neutron skin
thickness of $^{208}$Pb, namely, only measures the nucleon densities
from $r=R_{m}= 7.61 \pm 0.04$ fm to about $r=12$ fm. Thus, the
measurement of the $\Delta R_{np}$ would be substantially simplified
in hadron scattering experiments which have been hampered by the
strong absorption in the nuclear interior. We stress that, contrary
to the usual understanding, the nuclear surface properties are in
fact not well constrained by the nuclear mean-field models obtained
by fitting nuclear masses and charge radii. For instance, both the
SLy5 and NL3 interactions give $R_{m}=7.62$ fm, but they provide a
substantial difference in the $\Delta R_{np}$ amounting to $0.12$
fm. In other words, it is exactly the ambiguity of the nuclear
surface profile that leads to the large uncertainty of the $\Delta
R_{np}$, because the radial distribution function $f(r)$ relies on
the fourth power of distance $r$ according to Eq. (3), causing a
drastic amplification of the error as $r$ increases. Fig. 4(c)
illustrates the error accumulation of the $\Delta R_{np}$ in hadron
scattering experiments for different regions,
which is obtained by analyzing the data in Ref. \cite{JZ} combined
with the `tomoscan' method. The error accumulation at distance $r$ is defined as
the error generated by the region from the nuclear center to $r$. It
indicates that the error also primarily originates from the surface
structure. Therefore, the surface profiles must receive particular
attention and be measured with a much higher accuracy.

\section{Summary}\label{intro}\noindent
 We have developed alternative methods in the present
study to explore the neutron skin thickness $\Delta R_{np}$ of
$^{208}$Pb and density dependence of symmetry energy. The main
conclusions are summarized as follows. i) We have established a high
linear correlation between the $\Delta R_{np}$ and
$J-a_{\text{sym}}$ on the basis of widely different nuclear
energy-density functionals. Accordingly, an accurate
$J-a_{\text{sym}}$ value sets a significant constrain on the $\Delta
R_{np}$, which turns out to be a much more effective probe than the
parity-violating asymmetry $A_{\text{PV}}$ in the current PREX. ii)
The symmetry energy (coefficient) $a_{\text{sym}}$ of $^{208}$Pb was
extracted accurately with the experimental $\beta^{-}$-decay
energies of heavy odd-$A$ nuclei and with the experimental mass
differences. Given that the symmetry energy $J$ has been well
determined recently, the $\Delta R_{np}$ in $^{208}$Pb was thus
predicted to be $0.176\pm0.021$ fm robustly. This conclusion would
be significantly meaningful to discriminate between the models and
predictions relevant for the description of nuclear properties and
neutron stars. iii) With the above derived $J-a_{\text{sym}}$, the
values of the slope $L$ of the symmetry energy at the densities of
$\rho=0.16$ fm$^{-3}$ and $\rho=0.11$ fm$^{-3}$ which are of great
concern, are predicted to be $54\pm16$ MeV and $48\pm6$ MeV
respectively. These results, together with the $\Delta R_{np}$ of
$^{208}$Pb, can be applied to explore some intriguing problems in
nuclear astrophysics. In particular, the derived $a_{\text{sym}}$
and $S(\rho_{A})$ serve as important calibrations for a reliable
construction of new effective interactions in nuclear many-body
models. iv) The symmetry energy at suprasaturation densities up to
$\sim 2\rho_{0}$ was predicted to be not stiff.  v) With the firstly
proposed `tomoscan' method, we concluded that to obtain the $\Delta R_{np}$
one needs to only measure the nucleon densities in $^{208}$Pb
from $R_{m} =7.61\pm0.04$ fm as the densities in the range of $r<R_{m}$
have no contribution to the $\Delta R_{np}$. Thus, the measurement on the $\Delta R_{np}$ is
significantly simplified in hadron scattering experiments which have
been hampered by the strong absorption in the nuclear interior. Incidentally, the
`tomoscan' method could be employed to analyze the halo structure in exotic nuclei.
vi) It has been widely believed that the nuclear surface structure is
well constrained in nuclear energy-density functionals and in
experimental measurements. However, within the `tomoscan' concept,
we have showed that it is not true but a complete illusion. To grasp
the $\Delta R_{np}$, one must especially concentrate on the dilute
matter located in nuclear surface which results in the dominant
uncertainty.

\section*{Acknowledgment}\label{intro}\noindent
This work was supported by the National Natural Science Foundation
of China under Grants No. 11405223, No. 11175219, No. 10975190 and
No. 11275271, by the 973 Program of China under Grant No.
2013CB834405, by the Knowledge Innovation Project (KJCX2-EW-N01) of
Chinese Academy of Sciences, by the Funds for Creative Research
Groups of China under Grant No. 11321064, and by the Youth
Innovation Promotion Association of Chinese Academy of Sciences.

\end{document}